%% file: main.tex
\documentclass[twocolumn]{aastex63}
\pdfoutput=1 
\usepackage{amssymb}
\usepackage{amsmath,amstext}
\usepackage[T1]{fontenc}
\usepackage{apjfonts}
\usepackage{booktabs}
\usepackage{tabularx}
\usepackage[figure,figure*]{hypcap}
\usepackage{makecell}
\usepackage{multirow}
\usepackage{graphicx}

\newcommand{\intrv}[3]{\ensuremath{#1^{+#2}_{-#3}}}

\newcommand{\msun}{\ensuremath{M_{\odot}}}

\newcommand{\pduty}{\ensuremath{p_\mathrm{duty}}}

\newcommand{\Bayestar}{\texttt{BAYESTAR}}
\newcommand{\lalinference}{\texttt{LALInference}}

\newcommand{\change}[1]{#1}

\begin{document}

\newcommand{\nuaffil}{Center for Interdisciplinary Exploration and Research in Astrophysics (CIERA)
and
Department of Physics and Astronomy,
Northwestern University,
2145 Sheridan Road,
Evanston, IL 60208,
USA}

\title{Localization of Compact Binary Sources with Second Generation Gravitational-wave Interferometer Networks}

\author[0000-0002-1128-3662]{Chris Pankow}
\affiliation{\nuaffil}

\author[0000-0003-4690-2862]{Monica Rizzo}
\affiliation{\nuaffil}

\author{Kaushik Rao}
\affiliation{\nuaffil}

\author[0000-0003-3870-7215]{Christopher P L Berry}
\affiliation{\nuaffil}

\author[0000-0001-9236-5469]{Vassiliki Kalogera}
\affiliation{\nuaffil}

\begin{abstract}
GW170817 began gravitational-wave multimessenger astronomy. 
However, GW170817 will not be representative of detections in the coming years --- typical gravitational-wave sources will be closer the detection horizon, have larger localization regions, and (when present) will have correspondingly weaker electromagnetic emission.
In its design state, the gravitational-wave detector network in the mid-2020s will consist of up to five similar-sensitivity second-generation interferometers. 
The instantaneous sky-coverage by the full network is nearly isotropic, in contrast to the configuration during the first \change{three} observing runs. 
Along with the coverage of the sky, there are also commensurate increases in the average horizon for a given binary mass. 
We present a realistic set of localizations for binary neutron stars and neutron star--black hole binaries, incorporating intra-network duty cycles and selection effects on the astrophysical distributions. 
Based on the assumption of an $80\%$ duty cycle, and that two instruments observe a signal above the detection threshold, we anticipate a median of $28$ sq.\ deg.\ for binary neutron stars, and $50$--$120$ sq.\ deg.\ for neutron star--black hole (depending on the population assumed). 
These distributions have a wide spread, and the best localizations, even for networks with fewer instruments, will have localizations of $1$--$10$ sq.\ deg.\ range. The full five instrument network reduces localization regions to a few tens of degrees at worst.
\end{abstract}

\section{Introduction} \label{sec:intro}

The gravitational-wave (GW) localization of the binary neutron star (BNS) coalescence GW170817~\citep{2017PhRvL.119p1101A} led to the prompt discovery~\citep{2017Sci...358.1556C,2017ApJ...848L..16S,2017ApJ...848L..24V,2017Natur.551...64A,2017ApJ...848L..27T,2017Natur.551...67P,2017ApJ...850L...1L} and multi-wavelength observation~\citep{2017ApJ...848L..12A} of a host of electromagnetic (EM) emission from the aftermath of the merger. 
The localization and discovery was enabled by several factors, primarily the fortuitous proximity of the source. 
GW170817's distance~\citep[$40~\mathrm{Mpc}$;][]{2019PhRvX...9c1040A} was well within the sky- and orientation-averaged ranges of the LIGO Hanford (H) and Livingston (L) instruments~\citep{2015CQGra..32g4001L}, leading to the loudest signal detected by a GW network. 
Virgo~\citep[V;][]{2015CQGra..32b4001A} had recently completed upgrades towards a second generation design configuration and joined the run about a month or so prior to GW170817. 
This formed the first three-instrument network realized since 2010, and had obtained its first binary black hole (BBH) discovery three days earlier~\citep{2017PhRvL.119n1101A}, demonstrating the utility of a third instrument by reducing the HL only localization region size from $\sim 1200$ to $60$ sq. deg. 
\change{The configuration of the GW detector network is central to its localization ability.}

Potential multimessenger events like BNS and neutron star--black hole (NSBH) binary mergers depend on rapid localization for maximal payoff --- the kilonova associated with GW170817 may not have been identified as effectively if the full localization had taken several hours or days. 
Despite the numerous spectra~\citep{2017ApJ...848L..18N,2017Natur.551...75S,2017Sci...358.1574S,2017ApJ...848L..19C}, and extensive suite of photometry~\citep{2017ApJ...851L..21V}, the early rise time of the kilonova would have provided additional information~\citep{2018ApJ...855L..23A}. 
Other studies~\citep{2012ApJ...748..136C,2013IJMPD..2260011W,2015ASSP...40...51G,2016JCAP...11..056P,2016ExA....42..165C,2016MNRAS.459..121C,2017ApJ...840...88C} have explored the payoff for multimessenger astronomy when detection and localization is possible over various time scales, as well as demonstrated optimization techniques using the sky localization~\citep{2016ApJ...831..190H,2016PASA...33...50K,2017ApJ...846...62S,2018MNRAS.478..692C}. 
In addition to the sky location, distance information and identification of a host galaxy can aid follow-up~\citep{2013ApJ...767..124N,2014ApJ...784....8H,2016ApJ...820..136G}.
Identifying the host galaxy (or its galaxy cluster membership) is also of importance for measuring the Hubble constant~\citep{1986Natur.323..310S,2017Natur.551...85A,2018PhRvL.121b1303V,2019ApJ...871L..13F}.
We investigate the two- and three-dimensional localization potential of the GW network at design sensitivity.

GW170817 was a once-per-run event~\citep{2016arXiv161201471C}, even as the GW network progresses towards design sensitivity. 
Typically, BNSs would be found closer to the averaged detection range for the network, leading to weaker observed emission.
Since the GW localization region is dependent on the measured signal-to-noise ratio~\citep[SNR;][]{2011NJPh...13f9602F,2015ApJ...804..114B,2018MNRAS.479..601D}, signals further away will be on average less well localized than GW170817. 
Pairing both weaker EM emission and worse GW localization, the case for NSBH is more difficult to deal with: many NSBH detections may lie beyond the limiting magnitude of current telescopes, and localization regions will be larger.

Moreover, since the localization region size scales roughly with the mass of the binary~\citep{2018ApJ...854L..25P}, the distribution of masses within the population also shapes the distribution of localization precision.
With only \change{two} BNSs and no \change{confirmed} NSBH detected by GW networks~\citep{2019PhRvX...9c1040A,2020ApJ...892L...3A,2020ApJ...896L..44A}, the cosmic population and merger rate of these sources is uncertain. To fully understand the expected ability of a given GW network to localize, we must take into account the effects of the population~\citep{2012ApJ...757...55O,2011ApJ...741..103F,2019ApJ...876...18F,2002ApJ...572..407B,2004ASPC..312..393P,2006MNRAS.368.1742D,2008MNRAS.386..553I,2013MNRAS.428.3618C,2014LRR....17....3P,2015ApJ...806..263D,2017PASA...34...58E,2017ApJ...846..170T,2018MNRAS.479.4391M,2018PhRvL.120s1103K,2018MNRAS.474.2937C,2019MNRAS.482.2234G} on the region distribution. 
Particularly for NSBH, the wide range of masses will increase and widen the localization distribution~\citep{2018ApJ...854L..25P}.

Additionally, a network containing $5$ instruments will not always have all $5$ operating. 
In effect, various subnetworks will be \change{participating}, and these subnetworks have differing localization performance. 
Over an observing period, some localizations will be performed with only $2$ or $3$ detectors. 
This leads to larger localization regions.

The combination of network sensitivity, binary population models, and duty cycles are all crucial pieces to accurately describe the localization capabilities of the GW detector network in the next five years. 
In addition to analytical studies~\citep{2010PhRvD..81h2001W,2011CQGra..28l5023S,2011CQGra..28j5021F}, end-to-end simulations performed in anticipation of the first two observing runs were done with \Bayestar{} and \lalinference~\citep{2014ApJ...795..105S,2015ApJ...804..114B,2016ApJ...825..116F,2041-8205-829-1-L15,2018MNRAS.479..601D}. 
Other studies have addressed various facets of localizations with a $3$-fold or larger network at design sensitivity~\citep{2013ApJ...767..124N,2014ApJ...784..119R,2017CQGra..34q4003G,2018CQGra..35j5002F,2018ApJ...854L..25P}. 
In the next five years, it is expected that two additional large-scale interferometers will be operational~\citep{collaboration2013prospects}: the Japanese cryogenic interferometer KAGRA~\citep[K;][]{2013PhRvD..88d3007A}, and LIGO-India~\citep[I;][]{ligoindia}. 
We present a suite of simulated localizations with realistic populations of compact binaries, examining the capabilities of the full second-generation GW network, and analyzing the effects of different astrophysical mass and spin distributions.

Section~\ref{sec:bayes_loc} details the Bayesian approach to GW localization; Sec.~\ref{sec:net_and_pop} outlines source populations and GW detectors, and 
Sec.~\ref{sec:loc_results} describes the results for (sky and volume) localization, including the improvement for post-second generation heterogeneous networks. 
The interplay of the source population, localization, and potential EM follow up is explored in Sec.~\ref{sec:emfollow}. 
Finally, we discuss the implications of the results in Sec.~\ref{sec:discuss} and conclude in Sec.~\ref{sec:concl}.

\section{Bayesian Gravitational-Wave Localization} \label{sec:bayes_loc}

Information on source locations is encoded in the relative times, phases and amplitudes of GW signals observed across a network~\citep{2010PhRvD..81h2001W,2011NJPh...13f9602F,2011CQGra..28j5021F,2014PhRvD..89d2004G}; to extract localization information we analyze signals with Bayesian parameter-estimation algorithms.
These algorithms range from rapid minute timescales \citep[\Bayestar{};][]{2016PhRvD..93b4013S}, to intermediate hour timescales \citep[\texttt{RapidPE};][]{2015PhRvD..92b3002P}, and possibly multiple day timescales \citep[\lalinference{};][]{2015PhRvD..91d2003V}. 
All of these methods are capable of producing a joint posterior density for the location of the source in three-dimensions; a region on the sky and the distance to the source. \Bayestar{} uses an ansatz~\citep{2041-8205-829-1-L15} to determine an approximate distance posterior conditional on the sky position.
\texttt{RapidPE} and \lalinference{} also produce posteriors for some or all of the physical parameter space (masses, spins, etc.), hence the speed trade off.

In this work, we use \Bayestar, as it is unfeasible to assemble the required statistics for all the desired network configurations with other codes. 
While \Bayestar{} assumes only a single mass and spin configuration per event, \citet{2017ApJ...834..154P} showed that in the context of NSBH, that the orientation and location of the source did not significantly correlate or enhance the estimation of the physical properties of the system, and it is reasonable to assume that the converse is also true. 
Extensive studies have shown that \Bayestar{}  localization results are in good agreement with \lalinference{} results for BNS systems \citep{2014ApJ...795..105S,2015ApJ...804..114B,2016PhRvD..93b4013S}.

\section{Interferometer Networks and Source Populations} \label{sec:net_and_pop}

\begin{figure}
\includegraphics[width=\columnwidth]{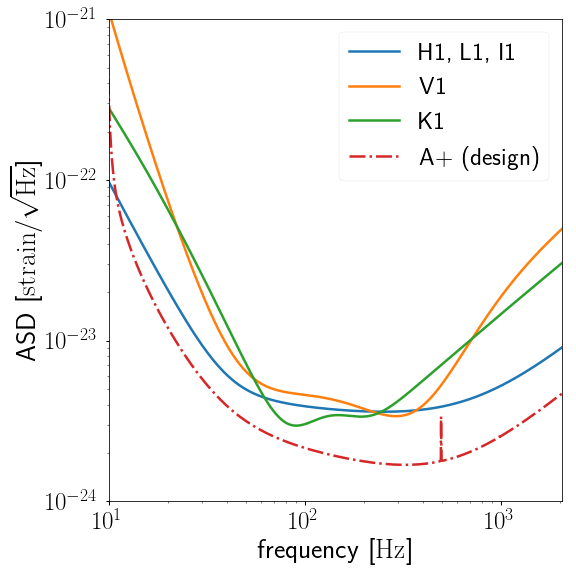}
\caption{\label{fig:net_sens} The strain amplitude spectral densities for the expected design sensitivity interferometers. The three LIGO instruments (Hanford, Livingston, and India) are anticipated to attain their design sensitivity (second generation), and one or more of the instruments may also be in a post-second generation (A+) configuration. KAGRA and Virgo, due to a differing instrumental set up, have slightly different spectral noise features, which resemble the design LIGO instrument sensitivity in the second generation.}
\end{figure}

Given the challenges in commissioning new detectors, it is difficult to predict sensitivity evolution. 
By the Collaboration's projections~\citep{collaboration2013prospects}, 2025 or later will see all $5$ instruments operating at the design sensitivity (the second generation curves in Fig.~\ref{fig:net_sens}). 
A proposed post-second generation configuration A+~\citep{LIGOT1800042} raises the question of a heterogeneous set of interferometers operating in tandem. 
Three interferometers, with sensitivities that differ over a factor of three, allowed for confident detections as well as enhanced sky localization for GW170814~\citep{2017PhRvL.119n1101A} and GW170817~\citep{2017PhRvL.119p1101A}. Therefore, it is essential to consider the entire network, and not only the most sensitive detectors, when considering localization ability.
We consider the set of interferometer configurations (for various duty cycles), with the instruments at design sensitivity, and we examine the consequences of a LIGO Hanford and Livingston A+ configuration. 
Recently, the anticipated sensitivity of the LIGO instruments was updated~\citep{LIGOT1800044,collaboration2013prospects}, reducing the overall detection range by a few tens of percent. 
This reduction in sensitivity would not drastically impact the conclusions reached here, shifting the overall distribution to larger values, but likely well within the uncertainties already associated with the simulation.

\change{The total number of instruments $N$ in a network will grow as additional instruments become operational.} 
\change{We distinguish between the number of detectors and the number which are taking data at the time of the observation as the \emph{participating} detectors $k \leq N$}.
For instance, a $3$-fold network configuration would have $N = 3$ instruments total but may only have $k = 2$ participating (perhaps the third is in maintenance at the time) for a given event. 

To consider a signal detected, we require that the SNR recorded in at least two instruments is above $5.5$.
While the SNR criterion is simplistic, it is roughly consistent with the $11$ events \change{from the first two observing runs}: the quietest events, GW151012 and GW170729, were found with SNRs of $\sim 9.5$--$10.8$ \citep{2019PhRvX...9c1040A} \change{corresponding to an average SNR per detector near our threshold}. 
The detection criteria for a GW event is typically not determined solely by the SNR, though it is a strong function thereof~\citep{2013PhRvD..88b4025C,2016CQGra..33u5004U,2016ApJ...833L...1A}. 
An analysis applying more sophisticated criteria would require a full search analysis with an extensive injection campaign with noise resembling the character of the search period. 
The threshold chosen here allows for a population of near threshold signals to be examined in addition to the near certain detection candidates; 
thus, we characterize the entire population of sources which are liable to be followed up with EM observations.

\subsection{Duty Cycles}

The interferometers are not in continuous operation during an observational run. 
Instruments are intentionally and unintentionally taken out of lock for a variety of reasons such as maintenance and environmental events (e.g., earthquakes, human activity), as such they can exhibit anthropomorphically induced cycles~\citep{2017ApJ...835...31C}. 
The duty cycle, like the sensitivity, is difficult to anticipate ahead of time.

To simulate the effect that varying duty cycles might have on events obtained during an observation run, we examine three cases: 
\begin{enumerate}
\item A $80\%$ duty cycle, which represents a value near the target operating point, high uptime with breaks allowed for maintenance and light commissioning.
\item A $50\%$ duty cycle, which may indicate degraded environmental conditions or unresolved technical issues with instrumental equipment.
\item A $20\%$ duty cycle, representing possible early commissioning phases \change{and engineering runs}.
\end{enumerate}
Different instruments can have differing and time-varying duty cycles. 
Most of the results presented here will still hold against minor variations on a fixed percentage uptime; however, given coordinated maintenance periods, as well as the previously mentioned cycles, it is likely that downtime between instruments will be correlated.

For a given duty cycle value, the probability of a network of $N$ total instruments operating with $k$ instruments \emph{participating} is proportional to the binomial distribution,
\begin{equation}
p(k \vert N, \pduty) = \binom{N}{k} \pduty^k (1-\pduty)^{N-k}. 
\end{equation}
For a duty cycle of $20\%$ ($\pduty=0.2$), the $5$-instrument network is effectively a set of $2$- and $3$-instrument networks, with a negligible probability of $4$ or $5$ simultaneously operating instruments. 
At the other extreme, at $80\%$ duty cycle only $< 1\%$ of the time has fewer than $2$ interferometers \change{participating} at any given time. 
We treat $k \leq 1$ as dead time: while \emph{detection} is possible~\citep[cf.][]{2020ApJ...892L...3A}, \emph{localization} is so broad (following the geometric sensitivity of a single interferometer) as to be unhelpful for follow up.

\subsection{Source Populations}

\change{The intrinsic physical parameters of the source, such as masses, affect localization, changing the localization for signals with similar strengths and sky positions. 
Once normalized by SNR, the localization region scales with the effective bandwidth (characterized by the noise-weighted Fourier moments of the signal); in turn, the bandwidth of the signal is inversely proportional to the chirp mass~\citep{2011NJPh...13f9602F,2011CQGra..28j5021F}. 
In terms of sky localization morphology, the annular regions are thinner for lower masses.} 

The physical parameter distributions of merging compact objects are not yet well measured. 
Population synthesis~\citep{2017ApJ...836L..26C,2018MNRAS.480.2011G,2018MNRAS.tmp.2091K} and empirical modeling~\citep{2003ApJ...584..985K,2008ApJ...672..479O,2017ApJ...846...82Z,2017ApJ...851L..25F,2018ApJ...856..173T,2019PhRvD.100d3012W,2019ApJ...870...71P,2019ApJ...876...18F} provide hints and limitations, but many of the inputs to binary evolution are still poorly constrained~\citep{2012ApJ...759...52D,2012ApJ...749...91F,2013A&ARv..21...59I,2017ApJ...836..244W}. 
As such we assume distributions where observational evidence suggest shapes~\citep{2019ApJ...882L..24A}, and take the widest possible distributions where they do not. 
Particularly in the BBH case, redshifts of greater than $1$ are achievable~\citep{2016ApJ...818L..22A}, and the complexities of determining rate and mass distributions with redshift~\citep{2018ApJ...863L..41F} are considerable. 

Orientation parameters, such as source sky direction, inclination, polarization angle, and coalescence phase are selected to correspond to uniform distributions, or isotropic in the case of spherical distributions. 
We distribute the sources in luminosity distance corresponding to redshifts uniform in the comoving volume out to the redshift horizon implied for the network under the SNR cuts applied. 
This distribution is supported by current observations of BBHs~\citep{2018ApJ...863L..41F,2019ApJ...882L..24A}. 

Since each source category is a mixture of different masses, the horizon in Table~\ref{tbl:param_dist} is calculated for the least asymmetric, most massive zero-spin configuration allowed by the population. 
This quantity is representative, since additional interferometers and certain aligned-spin configurations increase this number appreciably. 
The portion of the event populations localized here are filtered by detectability, with those not passing the SNR criteria rejected until a suitable sample size is obtained.

\begin{deluxetable*}{ccccc}
\tablecaption{\label{tbl:param_dist} Population parameter distributions. 
Spin directions are isotropic on the sphere. 
The redshift horizon is for a single interferometer at LIGO design sensitivity (blue line in Fig.~\ref{fig:net_sens}). 
Since this is a mass-dependent quantity, it is evaluated for a single mass and spin configuration: for the uniform distributions, the horizon is quoted for a maximum-mass, zero-spin binary, and for the two populations with Gaussian NS masses, we use the median NS  mass, and the maximum BH mass, assuming zero spin. 
\change{The detection horizon is thus close to the maximum for the population, but is not the absolute limit}.}
\tablehead{
\colhead{Population} & \colhead{BNS uniform} & \colhead{BNS normal} & \colhead{NSBH uniform} & \colhead{NSBH astro}
}
\startdata
\multirow{ 2}{*}{Mass distribution} &
\multirow{ 2}{*}{$m_{1,2} \in \textrm{U}(1, 2)$}  &
\multirow{ 2}{*}{$m_{1,2} \in \textrm{N}(1.33, 0.09)$} &
$m_1 \in \textrm{U}(3, 50)$ &
$m_1 \in \textrm{PL}(\alpha=-2.3, 3, 50)$ \\
 & & & $m_2 \in \textrm{U}(1, 2)$
 & $m_2 \in \textrm{N}(1.33, 0.09)$ \\
\hline
\multirow{ 2}{*}{Spin distribution} &
\multirow{ 2}{*}{$|\chi_{1,2}| \in \textrm{U}(0, 0.4)$} &
\multirow{ 2}{*}{$|\chi_{1,2}| \in \textrm{U}(0, 0.05)$} &
$|\chi_1| \in \textrm{U}(0, 0.99)$ &
$|\chi_1| \in \textrm{U}(0, 0.99)$ \\
 & & & $|\chi_2| \in \textrm{U}(0, 0.4)$
 & $|\chi_2| \in \textrm{U}(0, 0.4)$
 \\
\hline
Detection horizon redshift & 0.19  & 0.14  & 0.36  & 0.29 \\
(luminosity distance) & (980 Mpc) & (690 Mpc) & (2000 Mpc) & (1600 Mpc) \\
\enddata
\end{deluxetable*}

\subsubsection{Binary Neutron Stars} \label{sec:NS_params}

The bounds on the mass of a neutron star (NS) have not yet been exactly determined, but empirically, no NS with a mass smaller than $1.1 \msun$~\citep{2017ApJ...851L..29M,2018ApJ...854L..22S} has been confirmed. 
Masses much smaller than the Chandrasekhar bound are unlikely to exist given the processes which form NSs, though some processes such as ultra-stripped supernova are capable of producing such low mass NS~\citep{2015MNRAS.451.2123T,2017ApJ...846..170T}. 
The maximum mass is also yet undetermined. 
\change{EM measurements of BNSs~\citep{2018IAUS..337..146F,2017ApJ...846..170T} have not identified a NS heavier than $1.7 \msun$; the most massive Galactic pulsar is estimated to have a mass $1.9$--$2.1 \msun$~\citep{2020NatAs...4...72C,2020RNAAS...4...65F}; GW190425 is consistent with having a $1.2$--$2.5 \msun$ NS~\citep{2020ApJ...892L...3A}, and the potentially most massive known NS (in a binary with a main sequence companion) is $\gtrsim 2.5 \msun$ with significant uncertainties from orbital the inclination and rotation of the companion~\citep{2008ApJ...675..670F}}. 
Interpretation of GW170817~\citep{2018PhRvL.121p1101A,2020CQGra..37d5006A} has disfavored a stiff equation of state (EoS) which give rise to more massive maximum masses ($\gtrsim 2.5 \msun$) and tighter bounds have been inferred~\citep{2017ApJ...850L..19M,2020PhRvD.101f3007E}. 

The distribution of NS spins is also uncertain, observation provide some hints.
The fastest spinning NS in a BNS system is $\chi\sim0.05$~\citep{2003Natur.426..531B} depending on the EoS assumed, and the fastest known millisecond pulsar is spinning at $\chi\sim0.4$~\citep{2006Sci...311.1901H}. 

For the populations of BNSs, we consider two possibilities. 
The first is a broad distribution which intends to cover the widest available parameter space of merging BNSs: it is flat in a range of plausible masses $1$--$2 \msun$, and has dimensionless spin magnitudes up to $0.4$. 
The other is meant to emulate the Galactic population: masses following a Gaussian distribution with central mass $1.33 \msun$ and a standard deviation of $0.09~\msun$~\citep{2016ARA&A..54..401O}, with spin magnitudes up to $0.05$.

\subsection{Neutron Star Black Hole Binaries} \label{sec:BH_params}

As no NSBH have been confidently detected either by EM or GW instruments, even less is known about their intrinsic parameter distributions. 
High mass X-ray binaries (HMXRB) represent one possible path for formation --- Cygnus X-1, a $\sim 15 \msun$ main sequence star in a binary with a $\sim 10$ \msun black hole (BH) companion is a wind-fed HMXRB~\citep{2003ApJ...583..424G}. 
If the system survives the second supernova, it is possible that this system could form either a BBH or NSBH system, depending on supernova mass loss. 
Thus, it is not unreasonable to take these HMXRBs as examples. 
In \citet{2011ApJ...741..103F}, several models were fit to the distribution of the BH masses in HMXRB systems, and a power law was the most favored model, with an index of $\sim -4$. 
A similar analysis is presented for BBHs detected with GW instruments~\citep{2016PhRvX...6d1015A}. 
The GW BBH analysis obtains a power law index $\intrv{-1.6}{1.5}{1.7}$, with the result being correlated with the maximum BH mass~\citep{2019ApJ...882L..24A}. 
 
Measurements of HMXRB spins~\citep{2014SSRv..183..295M,2015ApJ...800...17F} are more challenging; a wide range of spins have been observed up to near maximal $\chi \sim 1$.

For NSBH, we again present results for two bracketing populations. 
One population is uniform and broad, taking on BH masses uniformly between $3$ and $50 \msun$, and BH spins up to near maximal, with NS spins up to $0.4$ (the reasoning for which is listed in Sec.~\ref{sec:NS_params}). 
This population covers the core-collapse supernova mass gap, a proposed depletion of BH between the most massive NS and $3 \msun$. 
Evidence for~\citep{0004-637X-741-2-103} and against~\citep{2012ApJ...757...36K} such a gap has been presented, and given the possibility of primordial~\citep{2016PhRvD..94h3504C}, and multi-generational mergers, we allow that this gap could be still be filled and emitting GWs. 
The second population has BHs distributed as a power law with index $-2.3$, which matches the slope of the initial mass function~\citep{1955ApJ...121..161S} and is compatible with the distribution for GW sources, and maximum mass $50 \msun$. 
The upper cut-off here is motivated by studies of the maximum BH mass and the putative \emph{second} BH gap induced by pair instability supernova~\citep{2017ApJ...836..244W,2019ApJ...882...36M,2019ApJ...887...53F}; such an upper mass gap is consistent with GW observations, which show a dearth of BHs with masses above $\sim 45 \msun$~\citep{2017ApJ...851L..25F,2019ApJ...882L..24A,2020arXiv200500023K}.

\subsection{Signal Model} \label{sec:sig_model}

We require a GW model to simulate the signals. 
In order to best capture the various features introduced by different source populations, we use the \texttt{IMRPhenomPv2} waveform family~\citep{2014PhRvL.113o1101H,2015PhRvD..91b4043S,2016PhRvD..93d4007K}.
While this family has been widely tested and is in use for observational property extraction~\cite{2019PhRvX...9c1040A,2019PhRvX...9a1001A}, there are cautions on the validity of the waveform for some spin and mass ratio configurations. 
\citet{2016PhRvD..94d4031S} showed specific regions of parameter space with pathological behavior. 
It is probable that the same or similar behavior is exhibited by the waveform family for some combinations of parameters \change{considered here}, particularly in the NSBH region where the mass ratio and spin configurations may exceed the limitations of the family. 
This manifests in \Bayestar{} with unphysical distance estimation and commensurately large $90\%$ credible regions which cover \change{the} entire sky. 
The unphysical scenarios are most easily identified in the inclination--sky area plane. 
We use machine-learning methods such as $k$-nearest neighbors to identify the cluster of spuriously localized events and remove them from the sample. 
There is a possible bias introduced from the possible misidentification of pathological results. 
If present, this bias likely decreases the upper end of the $90\%$ localization fractions quoted for NSBH in the tables throughout this work by a few tens of percent. 
\change{The \texttt{IMRPhenomPv2} model gives accurate results for the majority of systems we consider.}

\section{Localization Results} \label{sec:loc_results}

\begin{deluxetable*}{@{\extracolsep{4pt}}c cccc ccc}
\tablecaption{\label{tbl:skyloc_summ}Sky localization summarized by the median and the symmetric $90\%$ containment values of the $90\%$ credible regions $\Omega_{90\%}$ (in sq.\ deg.). 
The columns listed with duty cycles are computed from a set of $N$-fold distributions, weighted by the appropriate factors from the assumed duty cycle, see Eq.~\eqref{eqn:dcweight}.}
\tablehead{
\colhead{} & \multicolumn{4}{c}{Number of participating instruments $k$} & \multicolumn{3}{c}{Duty cycle $\pduty$} \\
\cline{2-5} \cline{6-8}
\colhead{Population} & \colhead{$2$} & \colhead{$3$} & \colhead{$4$} & \colhead{$5$} & \colhead{$20\%$} & \colhead{$50\%$} &\colhead{$80\%$}
}
\startdata
\input{2g_ndet_loc.tex}
\enddata
\end{deluxetable*}

A summary of the localizations for various combinations of networks, duty cycles, and populations is presented in Table~\ref{tbl:skyloc_summ}, where we quote the median and $90\%$ interval of the $90\%$ credible sky regions $\Omega_{90\%}$. 
The distribution for $N=2$ and $N=5$ is presented in Fig.~\ref{fig:bns_loc} for the two BNS populations and Fig.~\ref{fig:nsbh_loc} for the two NSBH populations.

The medians and intervals in Table~\ref{tbl:skyloc_summ} represent a wide range of potential localizations, but the progression of sensitivity with the number of detectors is clear. 
\change{There is a improvement of approximately five between $k=2$ to $k=3$.} 
When examining subnetworks (Sec.~\ref{sec:subnets}) of the $k=3$ and $k=4$ configurations, we find that there is no significantly \emph{better} network versus others combinations.

The results for $N=3$ compare well with previous work. 
\citet{2014ApJ...784..119R} considered a selection of BNS localized with the HLV and HILV networks at a fixed network SNR of $20$: their distributions for HLV are consistent with the BNS 3-fold (with all instruments above threshold) configuration here. 
The HILV results also match reasonably well with the 4-fold configuration, but their results are optimistic given their choice of fiducial SNR. 
The progression of HLV to HKLV to HIKLV for a set of uniformly distributed in mass NSBH events in \citet{2018ApJ...854L..25P} obtained similar values and improvements in localization region size.
\change{\citet{2018ApJ...854L..25P} obtained larger regions on the whole, but there is a likely bias that arises from artificially projecting a population of events detected in a $3$-instrument network into a $5$-instrument network. 
Different networks will observe a different set of events, particularly if they have instruments with differing spectral sensitivity shapes (as is the case for LIGO, Virgo and KAGRA).}

\subsection{Binary Neutron Stars} \label{sec:bns_loc}

\begin{figure*}
\includegraphics[width=3.5in]{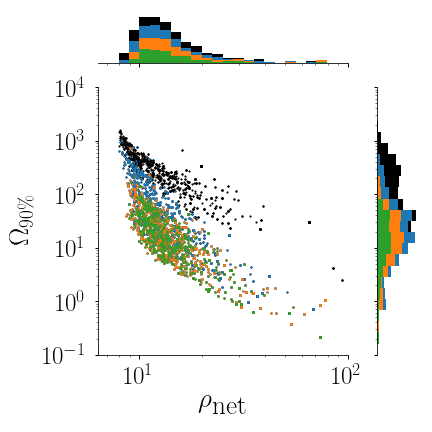}
\includegraphics[width=3.5in]{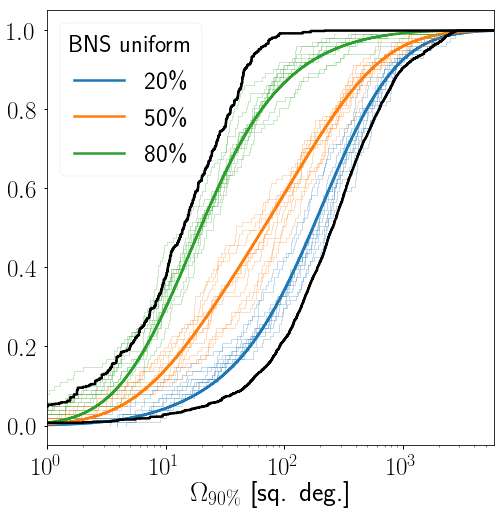}
\includegraphics[width=3.5in]{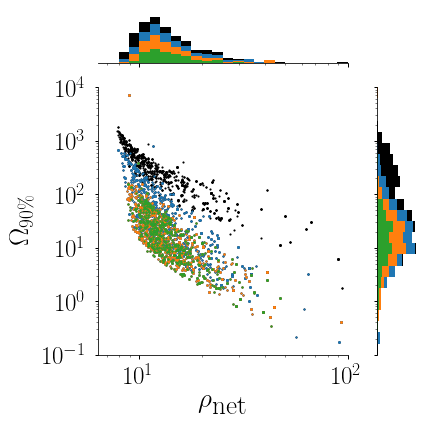}
\hspace{1mm}
\includegraphics[width=3.5in]{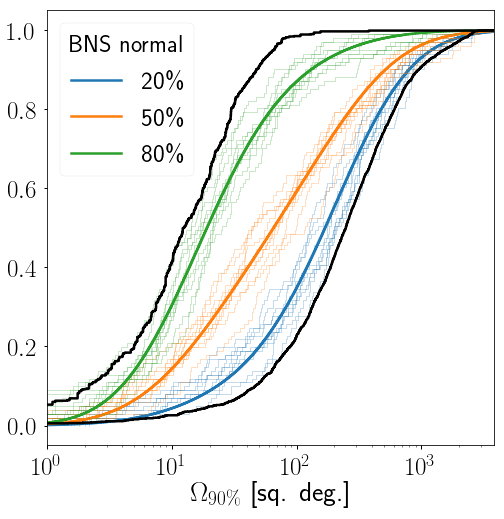}
\caption{\label{fig:bns_loc} The left column shows a scatter of the network SNR (abscissa axis) versus $90\%$ localization region (ordinate axis, sq.\ deg.) obtained for that event. 
Colors correspond to the number of \emph{total} instruments in the network --- black is two, blue is three, orange is four, and green is five. 
The stacked histograms of either axis are presented in marginalized histograms to the top and right of each scatter plot. 
The marginals are formed from downsampled versions of the overall sample since there is an uneven number of samples in each of the categories. 
The right column shows cumulative distributions of BNS localizations under the assumed models. 
The solid colored lines is the CDF of the fitted and weighted distributions constructed from Eq.~\eqref{eqn:dcweight}. 
The lighter step curves represent example realizations of $100$ events drawn from the overall sample and weighted appropriately by the duty cycle factors. 
The blue curves correspond to $20\%$ duty cycle, orange to $50\%$ duty cycle, and green to $80\%$ duty cycle. 
The two solid black lines bracketing those distributions are the full CDFs of all events in the $5$-instrument category (left black curve) and two-instrument category (right black curve). 
These represent best and worst case distributions. 
The light traces in the right panels correspond to a realization formed by drawing $100$ localizations from the $k$-fold configurations in proportion to the probability mass of the $k$-fold configuration given five total instruments to choose from. 
Top row corresponds to the uniform BNS distribution and the bottom is the normal distribution.}
\end{figure*}

Figure~\ref{fig:bns_loc} shows a summary of the localization distributions for the two BNS source types and networks. 
There is no statistically significant difference between the uniformly and normally distributed populations. 
Since most of the BNS in either population span the entire bandwidth of any of the instruments considered here, the localizations are expected to be similar, since one would obtain similar effective bandwidths~\citep{2011CQGra..28j5021F}. 
For example, a $2\msun+2 \msun$ binary's innermost stable orbit corresponds to a GW frequency of $\sim 1000~\mathrm{kHz}$, well outside the most sensitive frequencies of any of the three interferometer types in Fig.~\ref{fig:net_sens}.
The two populations differ in spin distributions, but the BNS spins are not expected to significantly influence localization \citep{2016ApJ...825..116F}, and this is the case here. 
BNS localization is effectively independent of details of the population, and current uncertainty in the astrophysical properties of BNS should not impact forecasts of localizations precision.

\change{When GWs travel over cosmological distances they become redshifted, and the merger appears to occur at lower frequency. 
This effect changes the effective bandwidth of a signal and increases the detected masses versus the source masses by a factor of $1+z$, where $z$ is the source redshift~\citep{1987GReGr..19.1163K}. 
Since BNSs are only detected at low redshifts, cosmological effects have a negligible effect on their localization properties, particularly relative to NSBH.}

\subsection{Neutron Star Black Hole Binaries} \label{sec:nsbh_loc}

\begin{figure*}
\includegraphics[width=3.5in]{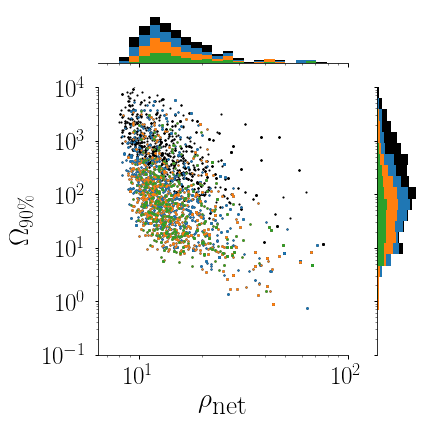}
\includegraphics[width=3.5in]{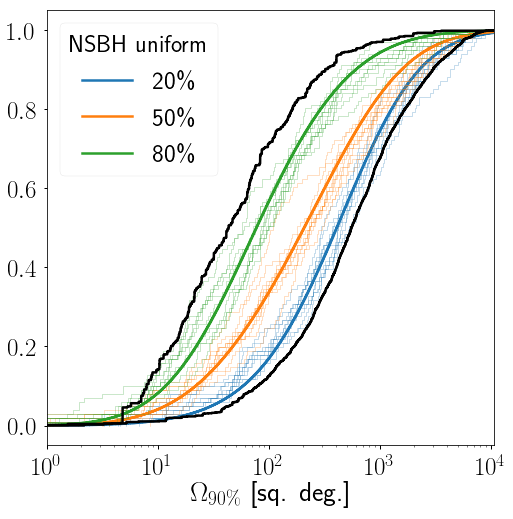}
\includegraphics[width=3.5in]{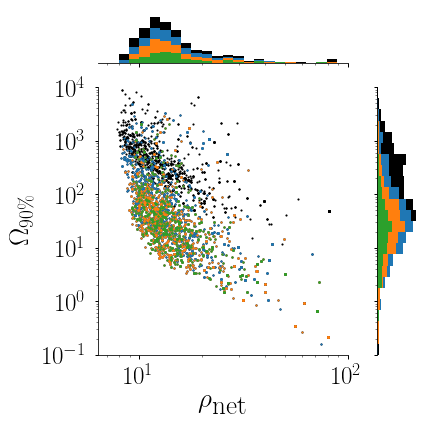}
\hspace{1mm}
\includegraphics[width=3.5in]{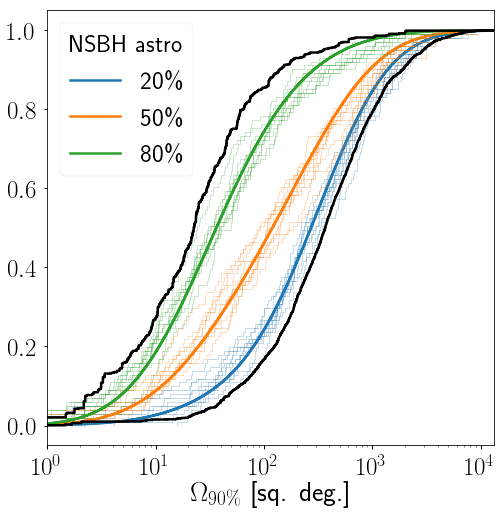}
\caption{\label{fig:nsbh_loc} Same as in Figure \ref{fig:bns_loc}, but for the uniform NSBH distribution (top row) and astrophysical NSBH distribution (bottom row).}
\end{figure*}

Figure~\ref{fig:nsbh_loc} summarizes the localization region distributions for the two model NSBH populations. 
In contrast to the BNS sets, the two NSBH distributions are significantly different. 
The astrophysical distribution is better localized by a factor of $1.5$ for $k=2$, and a factor of $2$ for $k=3,4,5$. 
This is a consequence of the astrophysical distribution containing more low mass binaries; these binaries have signals which extend to higher frequencies giving them greater effective bandwidths, and better sky localizations.
\change{In contrast, the uniform BH mass distribution contains more frequent high BH masses with smaller effective bandwidths, and thus a heavier tail of large localizations.}
The effects of the difference in mass distributions is compounded by cosmological effects. 
The most massive binaries are detectable out to the greatest distances, meaning that they suffer the most significant redshifting, which further decreases their effective bandwidth.
\change{Comparing the BNS and NSBH populations, there are more severe differences in effective bandwidth due to range of masses. 
This difference in source properties leads to a difference in localization distributions. 
Neglecting spin effects, the most direct comparison is between the BNS normal and NSBH astro sets, since the NS distribution is the same in both. 
For these, the typical NSBH total mass is $\sim1.5$--$2$ as heavy as for BNSs, and the median sky localizations differ by similar factors.}
The mass distribution of NSBHs does have noticeable consequences on our ability to localize the source.

\subsection{Duty Cycle Effects}

We consider here the effect of duty cycles on the expected localization distributions. 
The intervals reported in Table~\ref{tbl:skyloc_summ} for a given duty cycle are calculated by fitting each $N$-fold sample set to a log-normal distribution. 
Then each of the distributions are added together by weighting the contribute of each appropriately, so the distribution for a given sky localization accuracy $\Omega_{90\%}$ is given by
\begin{equation} \label{eqn:dcweight}
p(\Omega_{90\%} \vert N, \pduty) = \sum_{k\,=\,2}^{N} p(\Omega_{90\%}|k) p(k\vert N, \pduty) ,
\end{equation}
where $N=5$ is the total number of instruments and $k$ indicates the $k$-fold configuration. 
The cases $k=0,1$ are excluded explicitly, so the entire probability is renormalized after removing them. 
The relative weighting of each network according to its volumetric sensitivity is not accounted for --- we discuss the implications in Sec.~\ref{sec:discuss}. 
The right columns of Fig.~\ref{fig:bns_loc} (BNS) and Fig.~\ref{fig:nsbh_loc} (NSBH), show a selection of realizations for different duty cycles. 
The solid black lines to either side of the colored realizations represent a best and worst case scenario: they are the cumulative distributions for the $k=2$ (rightmost line), and the $k=5$ (leftmost line) configurations. 
The $5$-fold configuration implies an unrealistic $\pduty = 100\%$. 
Equally, the worst case scenario does not represent a physically realizable duty cycle, since for any $\pduty > 0$ will produce a non-zero set of times where $k > 2$. 
The duty cycle has a significant impact on localization accuracy, with the median $\Omega_{90\%}$ increasing by an order of magnitude between $\pduty = 20\%$ and $80\%$.

Even for an 80\% duty cycle, the performance of the network is not near optimal, the medians and intervals resemble the 4-fold network value, but with a wider spread. 
While the $k>3$ configurations do contribute about three quarters of the localizations, the $k=2,3$ configurations are the other quarter, \change{and those localizations are factors of several larger (hundreds of sq.\ deg.\ for $k=2$ versus a few sq.\ deg.\ for $k>3$ in Table~\ref{tbl:skyloc_summ}).}

\subsection{Distance and Volume Reconstructions}

\Bayestar{} is capable of providing a joint posterior on both sky location as well as distance. 
It does so by apply a per sky pixel ansatz on the distance posterior, assuming it is proportional to a Gaussian distribution weighted by a volumetric luminosity distance ($d_L^2$) prior~\citep{2041-8205-829-1-L15}.\footnote{The prior does not include adjustments to the luminosity distance from cosmological expansion.} 
Understanding the conditional distribution of distance on sky location is a useful tool; with a fiducial EM emission model it can provide limits on the source magnitude. 
This provides rapid answers to whether an instrument would realistically capture a source, or if a false positive is unnaturally bright and could therefore be discarded.

\begin{figure*}
\includegraphics[width=\linewidth]{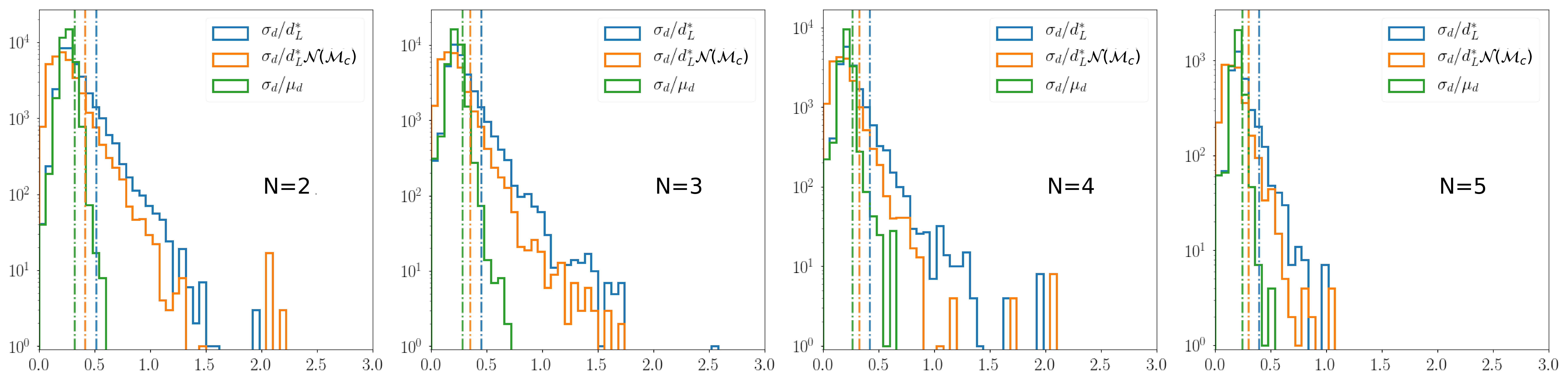}
\caption{\label{fig:dist_recon} Histograms of various measures of relative distance uncertainty. Blue histograms show the \Bayestar~estimated width divided by the true distance, orange show the same, but with an overall normalization which removes the first-order mass dependence on GW amplitude, and the green shows the width relative to the estimated mean of the marginal distance distribution estimated by \Bayestar.}
\end{figure*}

Following \citet{2015ApJ...804..114B}, we present the marginalized distance distribution standard deviations $\sigma_d$, normalized to the true distance to the source $d_L^*$ in Fig.~\ref{fig:dist_recon}, as well as the true distance with an additional normalization to remove the mass dependence. 
The mass normalization scales away the leading order dependence of the amplitude on the mass, specifically, we scale by the ratio $\mathcal{N} = \left(\mathcal{M}_{c,0} / \mathcal{M}_c\right)^{5/6}$, where $\mathcal{M}_c$ is the chirp mass of the binary and $\mathcal{M}_{c,0}$ is the chirp mass of a fiducial $1.4 \msun + 1.4 \msun$ BNS. 
Since we will not have either the distance or mass information known a priori, we also present $\sigma_d$ normalized by the reconstructed mean $\mu_d$ of the marginalized distance distribution.
\change{In all cases, values normalized by the mean, are more tightly constrained than the other two measures. 
This is because when the uncertainty is large, the long tail at large distances will pull $\mu_d$ to a higher value.}
Over all $k$-fold configurations and mass distributions, normalized distance uncertainties peak around $0.25$ with few events above $0.5$.
\change{With the $k=5$ network, the distance uncertainties become more consistent, with effectively no tail of events with $\sigma_d/d_L^* > 1$.}

\begin{figure}
\includegraphics[trim={0.5in 0.0in 0.8in 0.8in},clip,width=\columnwidth]{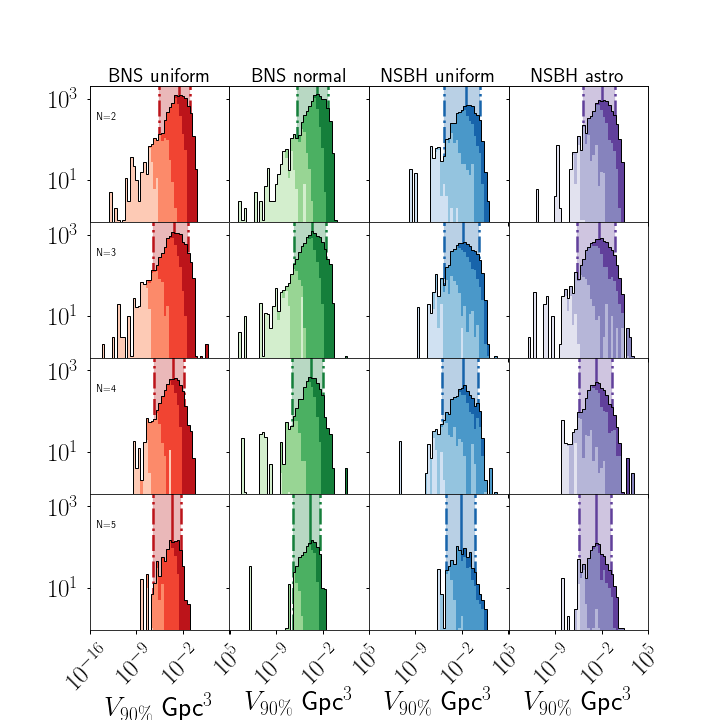}
\caption{\label{fig:vol_recon} For each of the four source populations, and $N$-fold networks, the $90\%$ posterior probability volumes are histogrammed, with shades of color indicating SNR bins. The darkest shade of each color are events with network SNR between $12$ to $20$, then becoming lighter for $20$ to $50$, and greater than $50$. 
From left to right: uniform distributed BNS are histogrammed in red shades, Gaussian BNS in green, uniform NSBH in blues, and the astrophysical NSBH mass distribution in purples. 
The medians and highest probability $90\%$ regions are shaded vertically in the background, with the median indicated by a solid line and the interval extremities indicated by a dot-dashed line of the appropriate color.}
\end{figure}

The volume localization will translate the number of galaxies which could potentially have been a given source's host~\citep{2014ApJ...784....8H}. 
This information is important for measurements of the Hubble constant~\citep{1986Natur.323..310S,2017Natur.551...85A}, as well as to give a rough idea of how many galaxies would need to be followed up to confidently observe any EM counterpart. 
Analogous to the $90\%$ credible area $\Omega_{90\%}$ for sky localization, we similarly define a $90\%$ credible volume $V_{90\%}$. Credible volumes for the various source populations are shown in Fig.~\ref{fig:vol_recon}.

Following the rows from top to bottom in Fig.~\ref{fig:vol_recon} shows the improvement in volume containment using networks with more instruments. 
\change{The volume localization depends upon the sky localization, the distance and the distance uncertainty~\citep{2018MNRAS.479..601D}. 
When considering different subnetworks, the greatest variation is in the sky localization, and this is the primary cause of variation in the volume localization.}
Gaussian distributed BNS have a $2$-fold median of $8\times 10^{-2}~\mathrm{Gpc}^3$, which \change{improves to $4\times 10^{-3}$ Gpc$^3$ with the 5-fold network}, similar gains are obtained for uniformly distributed BNS, but the medians are about twice as large, which reflects the more distant horizons achievable with higher mass binaries available in the uniform set. 
Increasing the number of instruments in the network also gives corresponding increases to the network SNR distribution. 
Hence the $5$-fold configuration has many more events (light shades) at correspondingly smaller volumes and higher network SNRs 
However, this effect is not very significant, increasing the median of the network SNR only by a unit between the $2$- and $5$-fold configurations.

\section{Subnetworks and Heterogeneous Networks} \label{sec:subnets}

\begin{figure*}
\includegraphics[width=\textwidth]{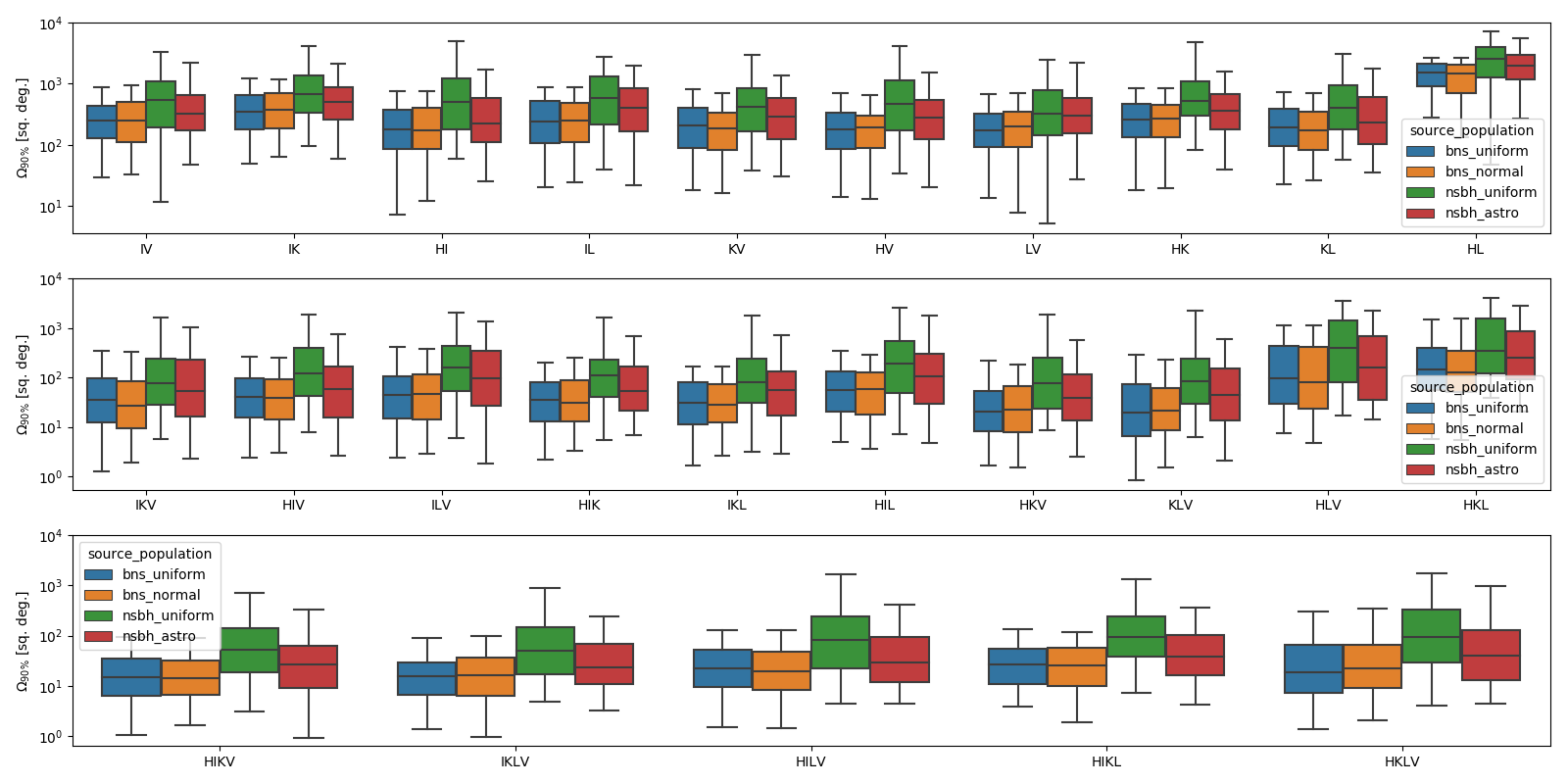}
\caption{\label{fig:subnet} Box plots representing the localization distributions for all unique subnetworks within a given $k$-fold configuration. 
Each set of box plots shows the $5$--$95\%$ percentiles with whiskers, and the box itself is the $25$--$75\%$ interquartile range, with the notch in the middle representing the median. 
The top row is $2$-fold, the middle $3$-fold, and the bottom is $4$-fold. 
The $5$-fold configuration is not shown, as it has only itself as a subnetwork.}
\end{figure*}

The localizations presented in Sec.~\ref{sec:loc_results} take a $k$-fold detector configuration as a whole, integrating together all of the subnetworks. 
We can break down the localization capability of each distinct instrument combination (hereafter subnetwork) within the $k$-fold set. The results for each $k$-fold configuration are presented in Fig.~\ref{fig:subnet}.

Geographic separation and differing sensitivities distinguish the localization capability of some subnetworks from others. 
One known correlation is in signal response between the H and L sites~\citep{2014ApJ...795..105S}. 
These two interferometers are the most closely spaced by angular separation on the surface of the Earth. 
This combination is the worst in terms of localization capability, with a factor of more than three in the medians over the next worst (IK). 
Performances of other $2$-fold subnetworks are generally better, with medians of a few hundred sq.\ deg. 
$3$-fold networks reduce the disparity, but subnetworks including the HL double still tend to obtain wider localization regions, with HIL, HKL, HLV all having medians near $100$--$200$ sq.\ deg: the others are below $100$ sq.\ deg., with the best median coming from KLV at a median of $30$ sq.\ deg. 
All of the $4$-fold networks perform similarly, with medians of $20$--$40$ sq.\ deg. 
The HKLV subnetwork stands out in the width of the distribution of credible regions. 
Where the other $4$-fold subnetworks have roughly similar means and widths, the HKLV network is shifted to larger credible regions; the upper $95\%$ percentile is $\sim 500$ sq.\ deg.\ in contrast to the the others which are typically about $100$--$150$ sq.\ deg. 
Given the relative performance of subnetworks containing the HL pair, if optimizing for localization ability, it makes sense to prioritize coincident observing for other pairs. 
For example, if possible, maintenance periods should be coordinated between H and V, rather than H and L, to maximize HV observing time. 
There is no clear variation across in localization ability across subnetworks for different astrophysical populations --- the distributions for different populations scale roughly between subnetworks.

\subsection{Heterogeneous Networks}

\begin{figure}
\includegraphics[width=\columnwidth]{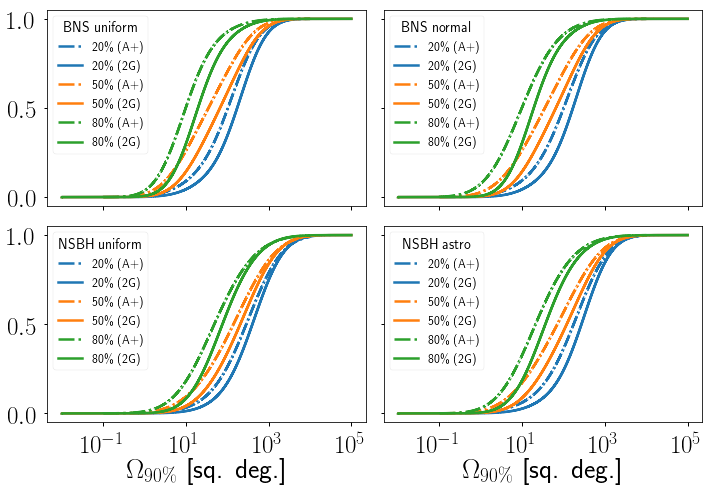}
\caption{\label{fig:aplus_loc_comp} Each panel shows the cumulative distribution function for its respective source type over both the second generation design networks (solid curves) and the mixed second-generation/A+ networks (dash-dotted curves). The duty cycles are colored as in other figures.}
\end{figure}

We also examine here the transition from a $5$-fold, design sensitivity network with second-generation instruments to a network with a heterogeneous set of instruments where the H and L instruments are upgraded to the A+ design. 
\change{We leave the investigation of upgraded versions of LIGO India, Virgo, and KAGRA to future studies, as the potential upgrade schedule is currently uncertain; see \cite{2018PhRvD..98b4029V,2018PhRvD..97j4064M} for investigations of the properties of events in the second- and third-generation of interferometers.}
\change{For this comparison, we use the same set of events from the design sensitivity study. This choice is to emphasize the impact of improved detector sensitivity relative to a baseline, and does not account for differences in the localization distribution of detected sources.} 
The overall shape of the localization distributions, see Figure~\ref{fig:aplus_loc_comp}, relative to their second-generation-only distributions remains mostly the same, but shifted to smaller localization regions.

\begin{deluxetable*}{@{\extracolsep{4pt}}c cccc ccc}
\tablecaption{\label{tbl:ap_skyloc_summ}Sky localization summarized by the median and the symmetric $90\%$ containment values of the $90\%$ credible regions $\Omega_{90\%}$ (in sq.\ deg.) for the A+ enhanced network. 
Column definitions are the same as those in Table~\ref{tbl:skyloc_summ}.}
\tablehead{
\colhead{} & \multicolumn{4}{c}{Number of participating instruments $k$} & \multicolumn{3}{c}{Duty cycle $\pduty$} \\
\cline{2-5} \cline{6-8}
\colhead{Population} & \colhead{$2$} & \colhead{$3$} & \colhead{$4$} & \colhead{$5$} & \colhead{$20\%$} & \colhead{$50\%$} &\colhead{$80\%$}
}
\startdata
\input{2g_aplus_loc.tex}
\enddata
\end{deluxetable*}

The improvement in the localization is enumerated in Table~\ref{tbl:ap_skyloc_summ}. 
\change{When the same events are localized with the design and A+ configurations, the localization distributions are uniformly improved, as expected for the boost in SNR.} 
All $k$-fold instrument networks, compared with Table~\ref{tbl:skyloc_summ}, see an overall $30$--$50\%$ improvement in the medians, and the spread in the credible regions decrease proportionally.

\change{
Breaking down the improvement via $2$-fold configurations, the overall improvement is not dominated by just contributions from the upgraded H and L.
The increase in sensitivity improves both the SNR and the ability of the network to do timing~\citep{2018CQGra..35j5002F}.
The HL configuration is the dominant detector pair by sensitivity, and thus will, in aggregate, contribute the most SNR when they are active.
However, as in Sec.~\ref{sec:subnets}, HL is not the network with the best localizations due to their relative geographical orientation. 
The enhancements to these instruments lead to narrowing in the width of the arcs but do not noticeably shorten the length of the arcs. 
}
There are significantly better localizations from other $2$-detector combinations. 
The improvement is best for the HL network versus any other subnetwork --- it sees significantly smaller regions, usually by a factor of $2$ or more; the other subnetworks involving H or L improve by typically less than a factor of $2$.

The volume distributions do not change appreciably in the bulk. For all configurations, the medians reduce by a factor of $2$, and the overall width of the distributions are reduced.

\section{Electromagnetic Follow-up Potential} \label{sec:emfollow}

Currently, the only GW signal to be confidently associated with an EM counterpart is GW170817~\citep{2017ApJ...848L..13A}.\footnote{A gamma-ray counterpart was associated with GW150914~\citep{2016ApJ...826L...6C}, but this statistical association is consistent with being by chance~\citep{2019ApJ...871...90B}.}
The GW event served as precursor to a host of emission processes across the EM spectrum, including a short gamma-ray burst~\citep[GRB;][]{2017ApJ...848L..12A} and r-process heating driven kilonova~\citep{1998ApJ...507L..59L,2017LRR....20....3M}. 
While both of these counterparts originated from the same merger, the emission properties are governed by significantly different post-merger mechanisms, and as such are moderated by different physical features of the pre- and post-merger objects. 
For GRBs, the probability of launching a jet has been phenomenologically linked~\citep{1993Natur.361..236M,2007NJPh....9...17L} to the presence of post-merger baryonic matter surrounding the system~\citep{2012PhRvD..86l4007F}. 
In the case of the kilonova, fits from numerical relativity~\citep{2016ApJ...825...52K,2017CQGra..34j5014D} simulations have provided a putative link between the properties of the inspiralling NS with the amount of dynamical ejecta contributing at least part of the kilonova medium. 
These fits neglect the role of disk winds~\citep{2015MNRAS.450.1777K,2017PhRvD..95f3016C,2019MNRAS.482.3373F}, which is an ongoing area of study. 
\change{To estimate whether a GW will have an EM counterpart, we must consider the availability of post-merger matter.}

\begin{figure*}
\includegraphics[width=\columnwidth]{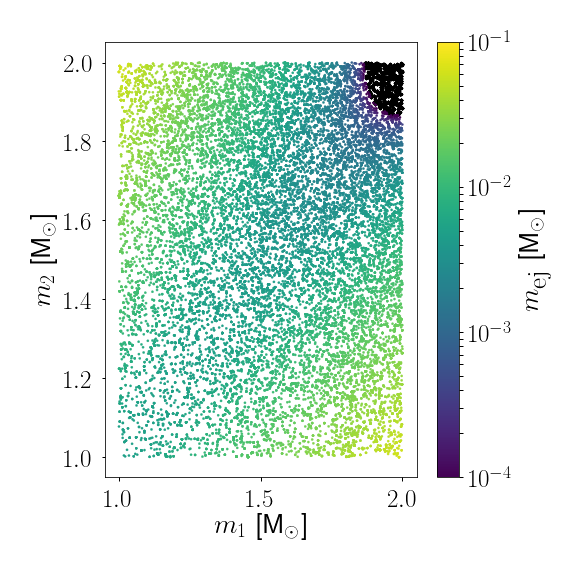}
\includegraphics[width=\columnwidth]{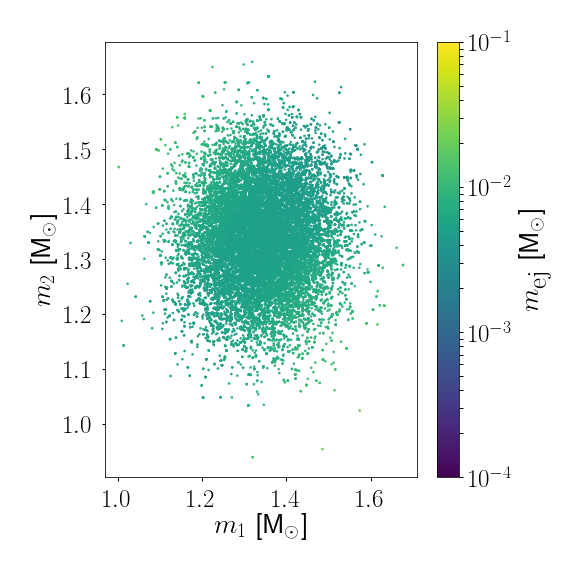} \\
\includegraphics[width=\columnwidth]{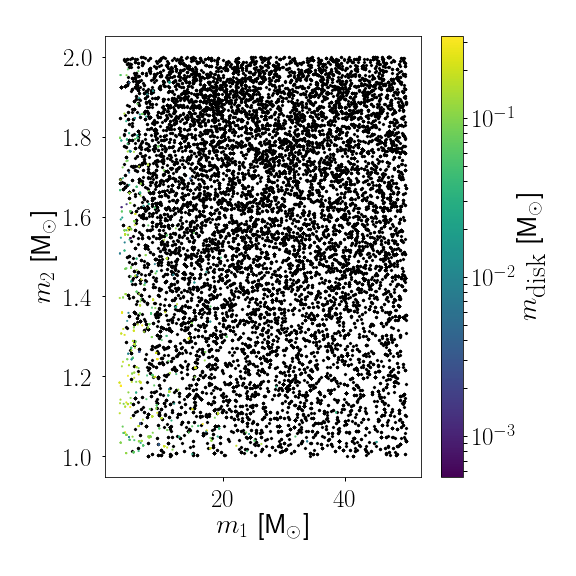}
\includegraphics[width=\columnwidth]{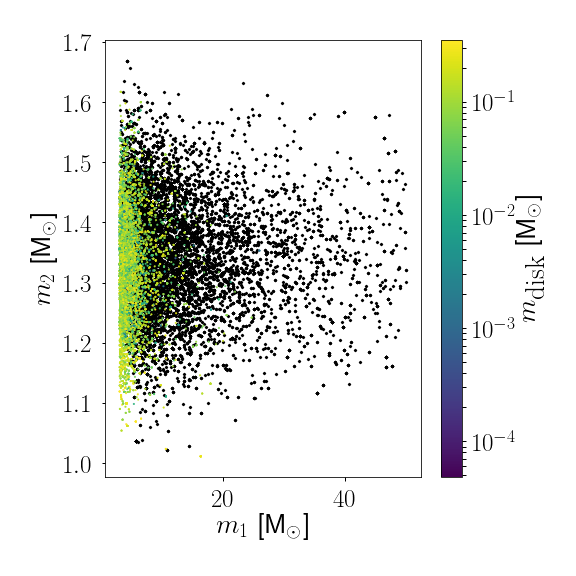}
\caption{\label{fig:em_bright} EM counterpart indicators for the four source categories, upper left and right correspond to projected dynamical ejecta mass~\citep{2017CQGra..34j5014D} over the component mass plane for BNS uniform and BNS normal respectively. 
The bottom row panels are the projected post-merger disk mass~\citep{2012PhRvD..86l4007F} for the uniform NSBH set (bottom left) and power-law NSBH set (bottom right). 
The color of the scatter indicates the expected ejecta mass, with black points implying that negligible ejecta would be produced to within the known uncertainties of the fits.}
\end{figure*}

The panels in Fig.~\ref{fig:em_bright}, represent simplified figures of merit for determining the amount of matter available to drive EM emission. 
For the BNS populations, we show only the projected dynamical ejecta mass distributions as no disk mass prescription is available. 
This impacts both our ability to predict a GRB as well as a component of the kilonova emission. 
We assume, however, that the presence of a kilonova implies a reasonable probability of enough matter to launch a GRB. 
Some caution is warranted in interpreting ejecta masses smaller than $\lesssim 10^{-3} \msun$, as the uncertainties in the fit would allow values consistent with zero.
From Fig.~\ref{fig:em_bright}, the amount of ejecta for BNS systems, is moderated both by the mass of the NS (more massive NSs have more ejecta) and by the mass ratio (more asymmetric systems produce more ejecta).
A less prominent effect is introduced by the EoS assumed to obtain the radius of the NS from its mass --- here we use APR4~\citep{1998PhRvC..58.1804A} which is a soft EoS whose maximum mass is not excluded by observation, and is consistent with current bounds on EoS from GW170817 itself~\citep{2018PhRvL.121p1101A,2020CQGra..37d5006A}. 
However, the results do not strongly depend on the choice of EoS, particularly those which are not excluded by observation. 
Fits for $2$- and $3$-component models of the ejecta from GW170817 produce a rough total of $\sim 5\times10^{-2}$ \msun~\citep{2017ApJ...848L..17C,2017ApJ...848L..18N,2017ApJ...848L..27T,2017Sci...358.1559K,2017ApJ...848L..19C,2017Sci...358.1570D,2017ApJ...851L..21V,2017Natur.551...75S}.
There is stark contrast between the uniform and normal populations of BNS; since the normal distribution is tightly concentrated it does not allow highly asymmetric and heavy NS required to produce significant amounts of ejecta. 

The bottom panels in Fig.~\ref{fig:em_bright} show the fitted disk mass from \citet{2012PhRvD..86l4007F}. 
Again, the difference in the mass distribution show definitive contrasts in the EM indicators across the NSBH mass space. 
The astrophysical power-law set tends to produce a higher fraction of events with non-negligible amounts of ejecta. 
Conversely to the BNS populations, where ejecta mass is enhanced by more asymmetric mass ratios, the NSBH populations favor \emph{less} symmetric combinations of component masses, whereas an unequal mass-ratio tends to suppress disruption of the NS and subsequently the available mass to form an accretion disk.\footnote{The BH spin also can allow for more asymmetric combinations.} 
Since the power-law distribution of BH masses concentrates most of the events towards more equal mass configurations, the fraction of positive EM indicators and distribution of ejecta mass is pushed to higher values relative to the uniform distribution where there is a much smaller fraction of systems potentially producing disks.

No appreciable correlation between localization area size and EM bright indicators is apparent. 
We tested this by checking the ejecta or disk mass distribution for events localized to $1$--$10$ sq.\ deg, $10$--$100$ sq.\ deg, and $> 100$ sq.\ deg. 
To within the the uncertainties from a finite sampling, no distinction between those categories is found when selecting for significant ejecta or disk masses.

\section{Discussion} \label{sec:discuss}

GW170817, detected with a network configuration which is closer to those tested in \citet{2014ApJ...795..105S} and \citet{2015ApJ...804..114B}, would be an outlier in those studies. 
They obtained few localizations with regions of the same size as GW170817. 
Considering the then-anticipated $3$-detector O2 results~\citep{2014ApJ...795..105S}, GW170817 falls within the top $\sim10\%$ in terms of $90\%$ credible region. 
GW170817 is exceptional on account of its high SNR which is a factor of $\sim2$ larger than that of the expected typical event~\citep{2011CQGra..28l5023S}. 
When viewed in the context of the $3$-fold column in Table~\ref{tbl:skyloc_summ}, GW170817's localization region is now more compatible with the median, though the obtained $90\%$ volume is comparatively small. 
In contrast to the distributions from earlier network configurations, that column summarizes networks whose overall reach has more than doubled with respect to the second observing run, so GW170817 returns a value around the median. 
Even at a duty cycle of $50\%$, GW170817's localization will be routine during that observing run. 
The estimated distances will often be estimated within $\lesssim 25\%$ accuracy, consistent with with modest improvement over networks examined in \citet{2015ApJ...804..114B}. 

The SNR threshold considered here ($5.5$ in two detectors) captures not only gold-plated detections, but also those which would be less significant. 
Another possible criteria is that the root-sum-squared SNR across the network $\rho_{\mathrm{net}}$ is above a given threshold. 
\cite{2015ApJ...804..114B} considered a threshold of $12$. 
Given the correlation of better localization with larger SNR, that the medians here are conservative; enforcing higher network SNR cuts will reduce the event count, but improve the distribution of the localization regions.

Our results neglect the impact of the relative volumetric sensitivity between networks --- surveyed volume translates directly into the mean detection count per observation time. 
At design sensitivity, Virgo surveys about $50\%$ less volume than the LIGO instruments, so networks with Virgo as the second most sensitive instrument will observe $50\%$ fewer events. 
This would diminish the relative contribution to $k=2$ with Virgo as the second most sensitive instrument. 
This disparity is most noticeable for the subnetworks including Virgo configurations --- the KAGRA and LIGO instruments have more similar surveyed volumes (KAGRA is $\sim 70\%$ of LIGO). 
For realistic duty cycles, events in $k=2$ containing only Virgo and another interferometer are rare enough as to not drastically affect the conclusions drawn here.

The localization algorithm used in this study assumes that the information on masses and spins provided are unbiased. 
For non-spinning sources, the extrinsic (orientation and location) parameters of the signal decouple almost entirely from the intrinsic mass parameters. 
Compact binary searches can measure the chirp mass of the system well~\citep{1993PhRvD..47.2198F,1994PhRvD..49.2658C,2016JPhCS.716a2031B,2019ApJ...884L..32B}, and since the dominant term in the post-Newtonian description of the waveform phasing is based on chirp mass, we do not expect any significant bias from non-spinning sources. 
Current GW binary searches~\citep{2016PhRvD..93l2003A} also incorporate the effect of spins aligned with the orbital angular momentum, and hence this information is also passed to \Bayestar --- our study assumes that the spin information is also perfectly measured. 
However, there is a degeneracy~\citep{1994PhRvD..49.2658C,2014PhRvD..89j4023C,2016ApJ...825..116F} between the mass ratio and the effective spin~\citep{2008PhRvD..78d4021R} which could lead to biases in reported mass and aligned spin components. 
Finally, searches do not incorporate the effects of spin tilts~\citep{1994PhRvD..49.6274A}, which have definitive imprints on the amplitude and phasing of the waveform. \Bayestar{} would then inherit any biases induced from this. 
To date, the BBH discovered so far have not shown large spins~\citep{2019PhRvX...9c1040A,2020arXiv200408342T}, but the NSBH in the population assumed here do have significant spin, anticipating the possibility. 
So, while the input populations themselves are unbiased, the compact binary searches and localization are probably suboptimal for a class of sources where the precessional impact is measurable.

Additional sources of uncertainty arise from the instrument noise. 
The sensitivities are representative, but we have assumed a zero noise scenario. \citet{2015ApJ...804..114B} showed that simulated signals injected into realistic instrument noise did not appreciable affect the outcome of the localization study. 
However, that study and this work ignore the effect of marginalizing over strain calibration uncertainty~\citep{2017PhRvD..95f2003A}. 
This will widen the localization and volume distributions presented here. However, the typical relative amplitude uncertainty is usually only a few percent~\citep{2017PhRvD..96j2001C}, and as such the widening is expected to not have a drastic effect on the distributions~\citep{collaboration2013prospects}.

The indicators of EM emission do not correlate strongly with the localization regions presented here. 
However, even if the localization performance is good, the outlook is not optimistic when population effects are accounted for. 
If the true BNS population resembles the Galactic one, then it is unlikely that many mergers will produce a large amount of dynamical ejecta, since this is driven by asymmetric masses. 
However, the fits for the Galactic BNS population have not yet been updated for newer (and more asymmetric) discoveries (a more up to date table can be found in \cite{2017ApJ...846..170T}), and there is evidence that GW170817 was also asymmetric~\citep{2018ApJ...866...60P}. 
\change{Furthermore, the discovery of GW190425~\citep{2020ApJ...892L...3A} shows that the observed Galactic population is not representative of all merging BNSs}. 
Taken together, the BNS population may be less like the Gaussian distribution and more like the uniform distribution where more asymmetric mergers are more common. 
The NSBH uniform distribution produces few ejecta products at all, since the extremely high mass ratios \emph{suppress} ejecta in this case --- no tidal disruption occurs and the NS is swallowed whole. 
The power law distribution of BH masses is more optimistic as the low end of the mass spectrum is favored. 
\change{Future multimessenger observations depend upon the (currently uncertain) underlying mass distribution as well as the GW network configuration.}

\section{Conclusions} \label{sec:concl}

This study has considered a realistic population of GW-detected BNSs and NSBHs. 
If the two BNS physical parameter distributions employed here could be considered bracketing, then the conclusion is that the variation over the mass and spin space does not appreciably affect localization area or volume. 
For NSBH, the distribution of localization regions is significantly affected by the mass distribution. 
When accounting for selection biases, the distribution of the masses (favoring less massive binaries) is less steep, because the detection volume scales strongly with the chirp mass. 
Since this favors more massive binaries, and they have intrinsically larger localizations, the $90\%$ region distribution is wider than what would be expected for a fixed fiducial $10 \msun+1.4 \msun$ system with randomized orientations and positions. 
As heavier systems are also found at typically larger redshifts, their redshifted signal resembles an even more massive binary, compounding this effect.
\change{Improved understanding of the NSBH mass distribution will enable more precise forecasts for localization.}

Our results imply that a relatively small fraction of signals will have EM signatures. 
Thus, considerable effort should be expended to maximize the duty cycles of each instrument in the network. 
A duty cycle of $50\%$ will both increase the median localization by \change{factors of four or more relative to $80\%$,} as well as induce a long tail of likely intractable sky localizations. 
Even $80\%$ is a factor of two away from the optimal $100\%$ performance. 
If a BNS is detected with a $3$-or-more-fold network, it should be localizable and with sufficiently fast and powerful telescopes, followed up. 
For instance \change{the Vera Rubin Observatory'}s~\citep{2008arXiv0805.2366I} or Zwicky Transient Facility's~\citep{2016PASP..128h4501B} native field of view should be able to tile most $3$-or-more fold skymaps in a single night without issue. 
NSBHs will be more challenging, being further away and subtending larger areas on the sky. 
Many of the sources should be localized spatially to within $\sim 10$--$25\%$ of their uncertainties scaled relative to their distance. Together, this implies that the closest and loudest BNSs will have volume reconstructions that will be tractable for galaxy-weighting schemes with good completeness within the local universe. 
Upgrading one or two of the instruments in the network to an anticipated post-second generation configuration brings a factor of two better localization area across all sources and duty cycles.

\acknowledgements
This work has benefited from discussions with and advice from Patricia Schmidt, Marica Branchesi, and Adam Miller. 
The authors are supported by the NSF grants PHY-1607709 and PHY-1912648, and the Center for Interdisciplinary Exploration and Research in Astrophysics (CIERA). 
This research was supported in part through the computational resources from the Grail computing cluster at Northwestern University --- funded through NSF PHY-1726951 --- and staff contributions provided for the Quest high performance computing facility at Northwestern University which is jointly supported by the Office of the Provost, the Office for Research, and Northwestern University Information Technology.

\software{
\texttt{numpy}~\citep{numpy}, \texttt{scipy}~\citep{2020SciPy-NMeth}, \texttt{sklearn}~\citep{scikit-learn}, \texttt{matplotlib}~\citep{Hunter:2007}, \texttt{seaborn}~\citep{michael_waskom_2020_3629446}, \texttt{lalsuite}~\citep{lalsuite}, \Bayestar{}~\citep{2016PhRvD..93b4013S}
}

\bibliography{references}

\end{document}

%% file: 2g_ndet_loc.tex
BNS uniform  &  \intrv{250}{1400}{230} & \intrv{42}{490}{40} & \intrv{19}{120}{18} & \intrv{14}{46}{13}  &  \intrv{180}{1100}{170} & \intrv{69}{820}{65} & \intrv{20}{240}{17} \\
BNS normal  &  \intrv{250}{1300}{230} & \intrv{40}{460}{37} & \intrv{19}{120}{17} & \intrv{13}{52}{12}  &  \intrv{170}{1000}{160} & \intrv{66}{790}{62} & \intrv{20}{230}{17} \\
NSBH uniform  &  \intrv{550}{4000}{500} & \intrv{130}{2500}{120} & \intrv{77}{1400}{73} & \intrv{43}{520}{37}  &  \intrv{410}{3300}{380} & \intrv{210}{2500}{200} & \intrv{78}{1000}{71} \\
NSBH astro  &  \intrv{370}{2600}{330} & \intrv{76}{1400}{72} & \intrv{32}{480}{29} & \intrv{22}{330}{20}  &  \intrv{280}{1900}{260} & \intrv{120}{1500}{110} & \intrv{37}{510}{33} \\

%% file: 2g_aplus_loc.tex
BNS uniform  &  \intrv{170}{570}{150} & \intrv{20}{180}{19} & \intrv{9.7}{50}{8.7} & \intrv{7.2}{21}{6.6}  &  \intrv{110}{620}{110} & \intrv{37}{470}{34} & \intrv{10}{120}{8.5} \\
BNS normal  &  \intrv{170}{550}{150} & \intrv{19}{190}{18} & \intrv{9.7}{50}{8.6} & \intrv{6.3}{25}{5.8}  &  \intrv{110}{620}{110} & \intrv{37}{470}{34} & \intrv{10}{120}{8.5} \\
NSBH uniform  &  \intrv{410}{3000}{390} & \intrv{89}{1800}{85} & \intrv{46}{1100}{42} & \intrv{27}{430}{23}  &  \intrv{310}{2500}{280} & \intrv{150}{1900}{140} & \intrv{49}{780}{45} \\
NSBH astro  &  \intrv{260}{1700}{240} & \intrv{40}{860}{38} & \intrv{18}{300}{15} & \intrv{13}{210}{11}  &  \intrv{190}{1300}{170} & \intrv{73}{970}{69} & \intrv{21}{300}{19} \\